# What Students Can Learn About Artificial Intelligence – Recommendations for K-12 Computing Education

Tilman Michaeli[1], Ralf Romeike[2] and Stefan Seegerer[2]

[1] Technical University of Munich, TUM School of Social Sciences and Technology, Computing Education Research Group
`tilman.michaeli@tum.de`
[2] Freie Universität Berlin, Computing Education Research Group
`ralf.romeike@fu-berlin.de`
`stefan.seegerer@fu-berlin.de`

**Abstract.** Technological advances in the context of digital transformation are the basis for rapid developments in the field of artificial intelligence (AI). Although AI is not a new topic in computer science (CS), recent developments are having an immense impact on everyday life and society. In consequence, everyone needs competencies to be able to adequately and competently analyze, discuss and help shape the impact, opportunities, and limits of artificial intelligence on their personal lives and our society. As a result, an increasing number of CS curricula are being extended to include the topic of AI. However, in order to integrate AI into existing CS curricula, what students can and should learn in the context of AI needs to be clarified. This has proven to be particularly difficult, considering that so far CS education research on central concepts and principles of AI lacks sufficient elaboration. Therefore, in this paper, we present a curriculum of learning objectives that addresses digital literacy and the societal perspective in particular. The learning objectives can be used to comprehensively design curricula, but also allow for analyzing current curricula and teaching materials and provide insights into the central concepts and corresponding competencies of AI.

## 1   Introduction

Artificial Intelligence is a central topic of computer science and has been a driving force of research and developments from the very beginning. In CS education, AI topics have always been an attractive way to motivate students to engage in the field of computing. As such, the development of games to play against the computer, or robotics, are common practice in CS education all around the world [1]. However, recent developments pushed the importance of AI forward significantly, attracting the attention of the media and making politicians require stakeholders in education to put a stronger emphasis on AI education. As a result, an increasing number of CS curricula are being extended to include the topic of AI. Furthermore, AI competencies are increasingly discussed as an important aspect of digital literacy for both teachers of all subjects (who need to understand the impact and application of AI technologies in their domain) and students (who experience a growing presence of AI technologies in their daily lives). A fundamental understanding of AI technology provides a key for two doors: the responsible use of such technology, and an informed discussion about the impact of AI on society.

Technological advances in the context of the digital transformation with increasingly powerful computing systems and a steadily growing volume of data are the basis for the rapid developments in the field of AI in the last years, especially in the field of machine learning. Consequently, whether computational thinking needs to be complemented by "AI thinking" [2] or as "CT2.0" [3] is a topic of discussion in CS education.

With longtime expertise in research on AI education and in developing teaching and learning materials for bringing the central concepts and ideas of AI to all levels of students, we frequently received requests by educational stakeholders for advice: What can and should be learned in the context of AI? Obviously, it is not sufficient to define content to be "taught"; it is necessary to define and discuss learning objectives that connect well to the established structures of educational systems and the understanding of teachers and students [4]. For AI, this remains a particular challenge since CS education research on central concepts and principles of AI is not yet sufficiently elaborated.

In order to address this challenge, in this paper, we present a curriculum of learning objectives suitable for mapping and understanding the field of AI education. In the following section, we discuss underlying goals and theories that such a curriculum needs to take into account with the goal of situating the topic and its challenges in the context of computing education. Section 3 highlights major developments in AI as well as related work on AI competencies in K-12 education. Section 4 presents the curriculum of learning objectives with a brief contextualization. The paper closes with a discussion of its applications and necessary future developments.

## 2  AI in the Context of CS Education

There is a consensus in computer science education research that teaching should focus on aspects fundamental to the subject and relevant in the long term. Short-lived technical developments, however, should play a lesser role. For this reason, various catalogs of principles, ideas, and concepts that characterize CS or one of its fields have been proposed over the past 30 years. These catalogs can be used, for example, in preparing new topics for teaching, as the foundation for curriculum development, and to provide insight into the field and its central aspects. According to [5], such characterizations also increase comprehensibility by shifting the focus from a technological perspective to underlying principles. They also enable achieving a "balance between concepts and practice" (ibid.) by highlighting the practices of the field and helping to provide a broader overview. Approaches such as the Fundamental Ideas of Computer Science according to Schwill [6], the Great Principles of Computing [7] or the Big Ideas of Computer Science according to Bell et al. [8] structure and characterize CS or its subfields by means of central terms, ideas, concepts, or underlying principles.

Since the field of AI is still undergoing rapid development with only little experiences and few studies on the integration of AI in education, we consider it important to start with the discussion of AI competencies and their contribution to general education. In order to be effective, such work needs to be put into perspective in a regional context and connect to the scientific and political discourse. In Germany, where this work originates from, a significant and helpful structure for understanding educational needs due to the digital transformation was achieved by the Dagstuhl Declaration [9]. Its stated objective is to enable students to use digital systems in a self-determined way. To this end, it is considered important to understand and explain digital systems, to evaluate them regarding their interaction with the individual and society, and to learn ways to use them creatively. Thus, for schools to fulfill their educational mission, phenomena, objects, or situations of the digital networked world should be viewed from three perspectives:

1. The technological perspective questions how digital systems work, explains their operating principles and teaches problem-solving strategies,
2. the socio-cultural perspective considers its interactions with individuals and society,
3. while the user-oriented perspective focuses on its effective and efficient use.

These equally important perspectives are referred to as the Dagstuhl triangle, which has also found its way into national education plans, e.g., in Switzerland. Considering these perspectives in the field of AI does not only connect well to the political discourse but may help assure that learning occurs based on a well-founded technological understanding, fostering applicability but also considering the significant impact AI has on society.

## 3 Developments in AI and Related Work

AI is the subfield of computer science that is concerned with replicating human cognitive abilities through computer systems and can be roughly divided into two major approaches: On the one hand, there are knowledge-based approaches to AI (sometimes also referred to as "classical" or "good old-fashioned" AI), which deal with the representation of knowledge and the drawing of conclusions through automated reasoning. Machine learning (ML) approaches, on the other hand, derive or identify rules, behaviors, or patterns themselves based on data - in other words, they "learn". This acquired knowledge is stored in a model and can subsequently be applied to new situations or new data.

AI problems are typically characterized either by a high degree of complexity or by the fact that they cannot be formalized conclusively, e.g. because of uncertainty. AI approaches build upon heuristics, probabilistics, statistics, planning, generalization, or reasoning that allow for dealing with these characteristics. However, typical AI systems are structured modularly and consist of multiple CS and/or AI tasks that work closely together. For example, speech recognition systems may involve aspects of hardware, software, pattern recognition, audio processing, and knowledge-based and ML approaches of AI.

Intending to develop guidelines for teaching AI to K-12, a working group identified five big ideas of AI [10]. These ideas comprise of the following:

1. Computers perceive the world using sensors.
2. Agents maintain models/representations of the world and use them for reasoning.
3. Computers can learn from data.
4. Making agents interact comfortably with humans is a substantial challenge for AI developers.
5. AI applications can impact society in both positive and negative ways.

So far, four of the ideas have been underpinned with concepts and learning objectives. However, the comprehensive list also includes learning objectives that are not specific to AI, such as how images or audio are represented digitally in a computer or illustrating how computer sensors work. Furthermore, discussions with stakeholders and teachers alike have revealed that they need a compact curriculum which serves their needs by connecting to both technical literature of the academic field (such as provided by [11]), as well as to established educational standards.

Another approach was chosen by Long and Magerko [12]. Based on an exploratory literature review of 150 documents such as books, conference articles, or university course outlines, the authors identified key concepts, which then formed the basis for their conceptualization of AI literacy – a set of competencies that everyone needs in the context of AI. They subdivide AI Literacy into five overarching themes in the form of questions: What is AI?; What can AI do? How does AI work?; How should AI be used?; How do people perceive AI? However, this approach is limited to a "historical" perspective, as it only reflects on existing material. This might be particularly problematic considering that CS education research on central concepts and principles of AI is not sufficiently elaborated yet. Furthermore, the competencies identified are on a rather general and abstract level.

In recent years, numerous methodological approaches have been developed for teaching AI in the classroom. They range from interactive experiments, unplugged-activities [13], configuring AI models/systems [14, 15], using models within programming projects [16-18] to implement AI algorithms [19-20]. All these approaches are mostly limited to a small set of particular competencies, but impressively illustrate the breadth of learning approaches for teaching a topic sometimes considered "too hard to understand".

## 4 Approach

This work was triggered by requests for a curriculum of learning objectives for the field of AI, which connects well with the recent political discourse, can be understood by teachers and stakeholders and takes the comprehensive experiences in CS education into account. It was started within a working group of experts from computing education research and practice. Over the course of one year, a list of AI learning objectives derived empirically as well as stated normatively with respect to the political discourse was curated, contrasted with technical literature and learning resources on AI, and compiled into a preliminary catalog of learning objectives. This catalog was then discussed with experts from the field of AI and K-12 CS education. The learning objectives were refined with the expert group once more, after incorporating this individual feedback, in an iterative process.

The developed catalog aims at two objectives:
(1) The curriculum of learning objectives should support the integration of AI as a topic in existing CS curricula. Thus, learning objectives typically already addressed in existing curricula are not included. Furthermore, learning objectives that are relevant in the field of CS, but not inherent to AI or AI systems, such as sensors or actuators, are omitted.
(2) The curriculum of learning objectives should underpin the importance of computing education as a basis for digital literacy and preparation for living in the digital world. To this end, learning objectives should not only focus e.g., on pure technological or pure societal aspects.

For educators and stakeholders of non-computing domains, this catalog may provide insight into the central concepts and corresponding competencies of AI.

## 5 Learning Objectives for Artificial Intelligence in Secondary Education

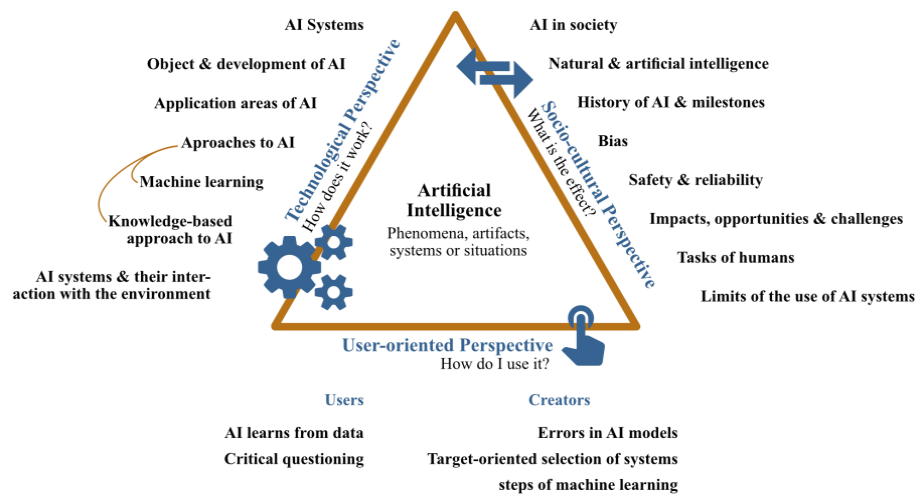

**Fig. 1.** Learning objectives for AI in secondary computing education according to the three perspectives provided by the Dagstuhl triangle

### 5.1 Technological Perspective (T)

The technological perspective provides insight into how phenomena, artifacts, and systems function or how they are structured. This is particularly difficult for the field of AI, considering that both the term and what is or is not considered AI are under a constant process of change. This ambiguity is reflected within common definitions of the field, such as "AI is the study of how to make computers do things at which, at the moment, people are better" [21]. Therefore, it is even more important that students are able to recognize AI systems in their daily lives, face different definitions and their implications, and characterize AI problems – in contrast to other problems in CS – and typical application areas. As the recent advancements in the field are primarily driven by advances in machine learning, there is a significant amount of attention on this particular domain of AI. However, other relevant approaches to AI must not be ignored – from a CS, as well as a computing education perspective.

Machine learning deals with algorithms that improve through experience over time [22]. Three different approaches to how machines can learn can be distinguished, which strongly depend on the overall goal and available data: supervised, unsupervised, and reinforcement learning. The corresponding competencies allow students to understand phenomena of their daily lives. For each of these approaches, there are various concrete methods that can make this idea of "learning" accessible within teaching. Crucial for machine learning and its success or failure is the available data which needs to be selected and preprocessed. As machine learning algorithms only learn from data (and therefore past experience), the difference between correlation (that those methods can identify) and actual causality is of utmost importance to assessing the capabilities and limits of such AI systems. The complexity of the models that are learned often provides a further challenge, as individual decisions of the system cannot be comprehended anymore. Therefore, understanding this loss of transparency and ways to tackle it (such as *explainable AI*) are crucial to enabling students to profoundly analyze the consequences of the usage of this technology.

Knowledge-based approaches to AI, however, are characterized by representing human knowledge in such a way that the computer can then be used for automatic reasoning. Often, knowledge-based and machine learning approaches are used together, supplementing each other. Given the goal of mimicking human intelligence, perceiving (using sensors) and interacting (using actuators) with the environment is a central task in many AI systems. Knowing that such systems typically have a modular structure and, for example, in working with language or images, consist of multiple computer science and/or AI tasks, is a core competency for understanding AI systems.

Students should be able to…

**T1 AI Systems**
...identify technologies that use AI methods.
...give indicators for when they are interacting with an AI system (e.g. with reference to the Turing test).
**T2 Object and Development of AI**
...discuss different definitions of AI,
...distinguish between strong and weak AI and give an example for each of these categories.
...distinguish between "AI problems" and other problems in computer science (e.g. with respect to uncertainty, direct relation of input and output) and describe approaches to deal with them (e.g. heuristics, probabilistics, statistics, planning, generalization, predictive and logical reasoning).
...explain the role of artificial intelligence in the history of computer science as well as the developments in computer science that have led to advances in the field (e.g. computing power, Big Data, "AI winter" and "AI summer").

**T3 Application Areas of AI**
...characterize application areas of artificial intelligence given their specifics (e.g. robotics, language processing, image processing, cognitive systems, artificial life).

**T4 Approaches to AI**
...distinguish and explain knowledge-based (sometimes also referred to as symbolic or "classical") and machine learning (sometimes also referred to as subsymbolic or data-driven) approaches to AI, state the fundamental differences between these approaches, and give typical examples of applications.

**T4.1 Machine Learning**
...describe different approaches to machine learning (reinforcement learning, supervised learning, unsupervised learning), explain their differences, and give examples of application in each case.
...assign concrete methods to the different approaches of machine learning and explain their basic functionality (e.g. k-nearest neighbors, decision tree learning, neural networks, linear regression, KMeans, vector quantization, Q-table learning).
...select the appropriate method in light of given data and goals.
...configure the hyperparameters (such as the number of neighbors for k-nearest neighbors) in suitable tools (e.g. in Orange).
...implement a concrete method to solve a problem.
...specify criteria to evaluate a trained model.

**T4.1a Data Selection and Preparation**
...decide which kind of data is needed for a given problem and prepare the data appropriately.
...explain why different design choices lead to different models.
...justify the procedure of dividing a data set into training and test data.
...describe how the training examples provided in an initial data set can affect the results of an algorithm.

**T4.1b Correlation and Causality**
...explain the difference between correlation and causality, give an example of each, and explain where these concepts are relevant in the field of AI.

**T4.1c Transparency and Explainability**
...distinguish transparency and explainability of AI systems.
...explain why the transparency of AI systems is often difficult to establish.
...name principles of algorithmic transparency and accountability.

**T 4.2 Knowledge-based Approach to AI**
...explain the approach and methods of knowledge-based AI approaches with reference to knowledge representation and reasoning.

...model knowledge explicitly in a representation form (e.g. as facts and rules, or semantic network, ...).

...explain different methods of reasoning (e.g. search, logical reasoning, probabilistic reasoning).

**T 5 AI systems and Their Interaction With the Environment**
...describe the modular structure of AI systems and divide an AI problem into different AI and computer science tasks.

...explain how AI systems collect data via sensors and interact with the world via actuators.

...illustrate that different sensors support different types of representation and thus give different insights about the world.

### 5.2 Socio-Cultural Perspective (S)

Within the socio-cultural perspective of the Dagstuhl triangle, the interactions of technology with individuals and society are addressed. Undoubtedly, AI severely affects society in many ways. The implications are vast but sometimes subtle. Therefore, it is all the more important that students are able to identify societal areas affected by AI. Furthermore, learning about AI also helps to strengthen understanding about humankind and natural intelligence in general and how the advancements of AI systems dovetail with the history of technology and society. As AI systems are increasingly incorporated into decision making, students have to be aware of the possibility of bias inherent in the data used for training and its influence on fairness and reliability of AI systems – once more reflecting other areas where e.g., problems resulting from representation issues are also common. As future shapers of society, students must be enabled to analyze the impact, opportunities, and challenges of AI. Furthermore, they have to know about ways to tackle potential problems of AI usage that help ensuring its responsible use. For this, it is crucial to clearly characterize the role humans are playing in creating AI systems. Only this way, an informed debate regarding the future of our society that takes opportunities as well as limits into account is possible.

Students should be able to…

**S1 AI in Society**
...identify and characterize areas of society affected by AI and find examples of AI from their daily life and classify them.

**S2 Natural and Artificial Intelligence**
...identify differences between artificially and naturally intelligent systems.

**S3 History of AI and Milestones**
...explain the history of AI and state milestones in its development and its importance to society (e.g. Deep Blue, Watson, AlphaGo, voice assistants).

**S4 Bias**

...explain why biases in data affect the results of machine learning and discuss implications for the use of AI systems.

**S5 Safety and Reliability of AI Systems**
...discuss the reliability of AI systems.
...name attack scenarios on AI systems (adversarial attacks) and classify them in terms of level (physical level, data level, protocol level).

**S6 Impacts, Opportunities and Challenges**
...analyze the implications, opportunities, and challenges of artificial intelligence for our society (e.g. the impact of automation on human workforce needs, idea of singularity, diversity, responsibility).
...explain ways to counter the problems (e.g. fake news in the context of deep fakes, analyzing and influencing human behavior) resulting from the use of AI (e.g. democratically determined fairness criteria, regulation of AI use, explainability).

**S7 Human Tasks**
...describe the tasks of humans when using AI systems (e.g. configuring, designing, critically assessing data).

**S8 Limits of the Use of AI Systems**
...explain the limits of the use of AI systems.
...explain misconceptions about the use of AI systems (e.g. Eliza effect, Tale-Spin effect and SimCity effect).

### 5.3 User-Oriented Perspective (U)

In the Dagstuhl triangle, the user-oriented perspective focuses on the purposeful selection of systems and their effective and efficient use. It includes questions about how and why tools are selected and used. For AI systems, we have to distinguish between two user scenarios within the user-oriented perspective:

(A) Consumers or end-users who use technology that passively incorporates AI (e.g. implicitly in apps, translation software, Alexa and co., self-driving cars) and

(B) Users who use AI actively for creating their own artifacts or solving AI problems by processing their own data sets – meaning they create, configure and use AI models explicitly (i.e. in systems such as Orange, MS Azure AI, LightSide or calling APIs such as Huggingface).

**(A) Consumer or End-User (non-creative)**

For consumers or end-users who use applications that have AI systems embedded, (especially reflective) competencies described in the socio-cultural perspective are sufficient. There are no AI-specific "operating skills", as AI is geared towards the user. However, since AI systems do have an enormous impact on our personal lives, it is all the more important to (based upon the technological perspective) be able to interpret and use the results provided by an AI system:

Students should be able to…

**UA-1 AI Learns from Data**
...explain that AI systems can learn from available data, including personal data, and make informed decisions regarding the disclosure of data in the interaction with AI systems.
...distinguish AI systems that apply generic AI models and AI systems that adapt to the user.

**UA-2 Critical Questioning**
...critically question the results of the conscious and unconscious use of AI systems (e.g. suggestions and prices in online stores).

**(B) Users who use AI actively for creating their own artifacts**

In contrast to mere end-users, users that employ AI in a "creative" manner by creating, configuring, and using AI models need to be familiar with AI methods and tools, but also be able to identify and correct possible underlying errors. In addition, respective tools must be chosen purposefully, and actual tool-specific "operating skills" are needed:

Students should be able to…

**UB-1 Errors in AI Models**
...explain why results of AI systems may contain errors, question obtained results, and identify and correct errors.

**UB-2 Target-Oriented Selection of Systems**
...name and justify selection criteria (e.g. with regard to data protection, bias, etc.) for deciding on an effective, efficient system in light of the data to be used and the goal.

**UB-3 Steps of Machine Learning**
...apply the steps of ML to solve a specific data-based problem (collect data, label as appropriate, select method, apply method, interpret results) in suitable tools.

## 6 Discussion and Outlook

Working with the catalog of learning objectives provided impressive insights into how much interested stakeholders and teachers can learn by simply reading it. Due to the recent omnipresence of machine learning, many are not aware of the breadth of the field of AI and the relevance of knowledge-based approaches. However, this part of AI has been very important in history, offers valuable learning experiences for understanding many AI approaches and may play a crucial role in hybrid approaches to AI which are getting increasingly important in AI research and development.

Furthermore, the socio-cultural perspective in particular is often overseen when new materials for AI topics are developed [23]. In line with the Dagstuhl triangle, we believe it is important that potentials and societal challenges are not discussed in isolation but based on a sound fundamental understanding of the technical fundamentals of AI.

An interesting question was to identify those learning objectives related to the use of AI systems: It is an inherent requirement of AI systems to be intuitive to use, removing the need for any special AI application skills. Similar to the proliferation of computer technology in the 1960s and 1970s, "demystification" is often seen as a primary educational goal, which can be achieved through the comprehension of basic methods [19]. Just as standard software that enables ordinary users to creatively implement their own problems and goals became increasingly important in the 1980s, we are also seeing more and more AI systems that enable ordinary users to evaluate their own data using AI methods to develop creative solutions. With the digital transformation and the digitization-related advancements of all disciplines, we believe that this aspect will become particularly important in schools in the future.

With the primary goal of answering the question of which distinct AI learning objectives might be important in secondary education, the question of an order and the relative importance of competencies, as well as that of the competency levels, was not considered in this work. Considering the currently very heterogeneous attempts to integrate the topic of AI into the curricula, this approach does not seem to be purposeful to us. Thus, the presented catalog can be used to comprehensively design half a year of CS lessons, but also to give students a brief insight into the topic. Furthermore, it allows for analyzing what competencies are addressed in current curricula and teaching materials and provides insight into the central concepts and corresponding competencies of AI – even for stakeholders of non-computing domains.

With the approach taken, it is obvious that the catalog does not include competencies on detailed levels, as i.e., AI4K12 [10] are slowly progressing to. However, the abstraction level chosen allows for quickly grasping possible focus areas of AI education and allows other disciplines to connect e.g., ICT/Media education to the application-oriented perspective or the humanities with the societal perspective.

Eventually, AI competencies must be merged into CS curricula. To this end, in a first step, what can and should be learned in the context of AI needs to be characterized. With this catalog of learning objectives, we present a recommendation to address this need, which has already proven to be helpful for several stakeholders in creating CS curricula.